\documentclass[twocolumn]{aastex62}
\usepackage{apjfonts,amsmath,verbatim}

\shorttitle{GW190521's Dynamical Formation and Prospects for Decihertz GW Astronomy}
\shortauthors{Holgado et al.}

\graphicspath{{./figures/}}

\begin{document}

\title{DYNAMICAL FORMATION SCENARIOS FOR GW190521 AND PROSPECTS FOR \\ DECIHERTZ GRAVITATIONAL-WAVE ASTRONOMY WITH GW190521-LIKE BINARIES}

\correspondingauthor{A.~Miguel Holgado}
\email{mholgado@andrew.cmu.edu}

\author[0000-0003-4143-8132]{A.~Miguel Holgado}

\author{Alexis Ortega}

\author[0000-0003-4175-8881]{Carl L.~Rodriguez}
\affil{McWilliams Center for Cosmology and Department of Physics, Carnegie Mellon University, Pittsburgh, PA, 15213
}



\begin{abstract}
The gravitational-wave (GW) detection of GW190521 has provided new insights on the mass distribution of black holes and new constraints for astrophysical formation channels. 
With independent claims of GW190521 having significant pre-merger eccentricity, we investigate what this implies for GW190521-like binaries that form dynamically. 
The Laser Interferometer Space Antenna (LISA) will also be sensitive to GW190521-like binaries if they are circular from an isolated formation channel. 
However, GW190521-like binaries that form dynamically may skip the LISA band entirely. 
To this end, we simulate GW190521 analogues that dynamically form via post-Newtonian binary-single scattering.
From these scattering experiments, we find that GW190521-like binaries may enter the LIGO-Virgo band with significant eccentricity as suggested by recent studies, though well below an eccentricity of $e_{\rm 10Hz} \lesssim 0.7$. 
Eccentric GW190521-like binaries further motivate the astrophysical science case for a decihertz GW observatory, such as the kilometer-scale version of the Midband Atomic Gravitational-wave Interferometric Sensor (MAGIS).
We carry out a Fisher analysis to estimate how well the eccentricity of GW190521-like binaries can be constrained with such a decihertz detector.
These eccentricity constraints would also provide additional insights into the possible environments that GW190521-like binaries form in. 
\end{abstract}


\section{Introduction} \label{sec:intro}
The gravitational-wave (GW) detection of GW190521 from the LIGO Scientific Collaboration and Virgo Collaboration (LVC) is the most black-hole (BH) merger observed so far \citep{ligo_scientific_collaboration_and_virgo_collaboration_gw190521_2020}.
This event marks a new milestone for GW astrophysics by revealing new insights into the mass distribution of BHs \citep{abbott_properties_2020}. 
The unusually high masses of GW190521, whose primary component lies within the ``upper mass gap'' of BHs, strongly suggest that the binary formed via dynamical encounters in a dense stellar environment \cite[e.g.,][]{2020ApJ...900L..13A,2020arXiv201105332K, 2020ApJ...902L..26F,2020arXiv200910068L,2020arXiv201102507G,2020ApJ...903..133S,2020ApJ...903...45K,2020ApJ...904L..13R}, where the hierarchical mergers of BHs or stars can produce objects more massive than those formed from the collapse of isolated stars.  
\par
While the presence of a BH in the mass gap is strong evidence for a dynamical formation scenario, it is not conclusive.
It is possible (i.e., not ruled out) that GW190521 could have formed from isolated stellar binaries \citep{belczynski_most_2020,costa_formation_2020,renzo_stellar_2020,tanikawa_population_2020} or within gas-rich environments \citep[e.g.,][]{roupas_generation_2019,rice_growth_2020,safarzadeh_formation_2020,toubiana_detectable_2020} where the GW signal itself may be affected by the accretion and external torques \citep[e.g.,][]{barausse_can_2014,holgado_gravitational_2019,caputo_gravitational-wave_2020}.  BH masses, however, are not the only possible indicator of a dynamical formation scenario.
In particular, two independent studies from \cite{romero-shaw_gw190521_2020} and \cite{gayathri_gw190521_2020} have found that GW190521 is  consistent with the binary having a significant amount of eccentricity as it entered the LIGO band, which has long been seen as a tell-tale sign of dynamical formation \cite[e.g.,][]{wen_eccentricity_2003,2012ApJ...757...27A,2014ApJ...781...45A,2019ApJ...871...91Z,2017ApJ...840L..14S,2014ApJ...784...71S,2018arXiv180506458A,2018ApJ...856..140H,2017ApJ...836...39S,2016ApJ...828...77V,gondan_eccentric_2018,michaely_high_2020,tagawa_eccentric_2020}.  
\par
In this Letter, we consider the astrophysical implications of an eccentric GW190521, with a particular emphasis on multiband GW astronomy \cite[e.g.,][]{amaro-seoane_detection_2010,sesana_prospects_2016}.  
We first explore the implications of GW190521's possible eccentricity for dynamical formation channels, particularly GW-driven capture during encounters of 2 or 3 BHs. 
Either of these processes can occur in globular clusters \citep[e.g.,][]{rodriguez_binary_2015,rodriguez_binary_2016,dorazio_black_2018} or nuclear clusters \citep[e.g.,][]{gondan_eccentric_2018,tagawa_eccentric_2020} and can be efficient in forming the stellar-mass BH binaries that the LVC observes. 
\par
GW190521-like binaries may also be sources for the Laser Interferometer Space Antenna (LISA), which will open up the millihertz band of the GW spectrum in the 2030s \citep{amaro-seoane_laser_2017}.
If such binaries are circular, LISA could be able to detect their wide inspirals before they eventually merge in the LIGO band \citep{sesana_prospects_2016,toubiana_detectable_2020}. 
If GW190521-like binaries form dynamically, however, they may skip the LISA band entirely and thus prevent a pre-merger GW observation at millihertz GW frequencies. 
We thus investigate the prospects for decihertz GW astronomy with GW190521-like binaries and estimate how well the eccentricity can be constrained before such binaries provide energy to the LIGO band.
\section{GW Captures during Two-Body Encounters}
Within dense stellar clusters, close encounters may occur among heavier stellar-mass compact objects that sink towards the center due to dynamical friction. 
Gravitational radiation during a close encounter may result in a capture, i.e., the energy of the binary transitions from positive to negative.
Successful captures can form highly eccentric binaries that then inspiral via GWs, decreasing both the semi-major axis and eccentricity towards merger. 
The eccentricity that remains as the binary enters the LIGO band can then be used to infer what the conditions were for a GW capture scenario in a dense star cluster. 
\par
We thus consider the distance of closest approach, the periapsis, for a close encounter between two unbound stellar-mass BHs with component masses $m_1$ and $m_2$. 
By equating the kinetic energy of parabolic encounters to the energy radiated in GWs in the quadrupolar approximation, one can estimate the maximum periapsis $r_{\rm p,max}$ required for GW capture \citep[e.g.,][]{quinlan_collapse_1987,berry_gravitational_2010} as
\begin{equation} \label{eq:rp}
r_{\rm p,max} = \left(\frac{85 \pi}{2^{3/2}{\cdot}3}\right)^{2/7} \frac{G(m_1 m_2)^{2/7} M^{3/7}}{c^{10/7}v^{4/7}} \ ,
\end{equation}
where $G$ is the Newton's constant, $c$ is the speed of light, $M$ is the total mass, and $v$ is the velocity of the encounter.
Any periastron distances above this maximum value will not result in a GW capture.
%
%
\par
With the possibility of finite eccentricity for GW190521 as it entered the LIGO band, we can estimate the semi-major axis $a$ and eccentricity $e$ at lower GW frequencies with the quadrupole approximation \citep{peters_gravitational_1963,peters_gravitational_1964}. 
Since GWs radiate away both orbital energy and angular momentum, both the semi-major axis and eccentricity decrease towards zero as GW inspiral proceeds. 
From the quadrupole approximation, the orbital frequency and eccentricity evolve as \citep[e.g.,][]{huerta_detection_2015,dorazio_black_2018},
\begin{equation}
\frac{f_{\rm orb}}{f_0} = \left[\frac{1 - e_0^2}{1-e^2}\left(\frac{e}{e_0}\right)^{12/19}\left(\frac{1+\frac{121}{304}e^2}{1+\frac{121}{304}e_0^2}\right)^{870/2299}\right]^{-3/2}\ .
\end{equation}
Given an eccentricity at a reference frequency, one can estimate the eccentricity at either higher or lower frequencies. 
An eccentric binary will emit over several harmonics, such that the peak harmonic primarily determines the GW frequency of the emitted waves.
One can estimate the rest-frame GW frequency of an eccentric binary using the following fitting formula \citep{wen_eccentricity_2003} 
\begin{equation} \label{eq:rest}
f_{\rm GW,r} = \frac{\sqrt{GM}}{\pi} \frac{(1+e)^{1.1954}}{\left[a\left(1-e^2\right)\right]^{3/2}} \ ,
\end{equation}
which is also related to the observed GW frequency as $f_{\rm GW,r} = (1+z) f_{\rm GW}$. 
\par
The rest-frame GW frequency can then be used to determine the binary semi-major axis $a$. 
Combining $a$ and $e$ can then be used to obtain the periapsis at formation
\begin{equation} \label{eq:form}
r_{\rm p,0} = a_0 \left(1 - e_0\right) \ .
\end{equation}
With Eqs.~\eqref{eq:rp} and \eqref{eq:form}, we can then determine what local velocity dispersion $\sigma$ is required in order to achieve GW capture. 
Dense star clusters will have a range of velocity dispersions that depends on the distance away from the cluster center. 
At any given location in the cluster, one can describe the local velocity distribution with a Maxwellian distribution, which we use when evaluating \autoref{eq:rp}. 
\par
\begin{figure}
\centering
\includegraphics[width=\columnwidth]{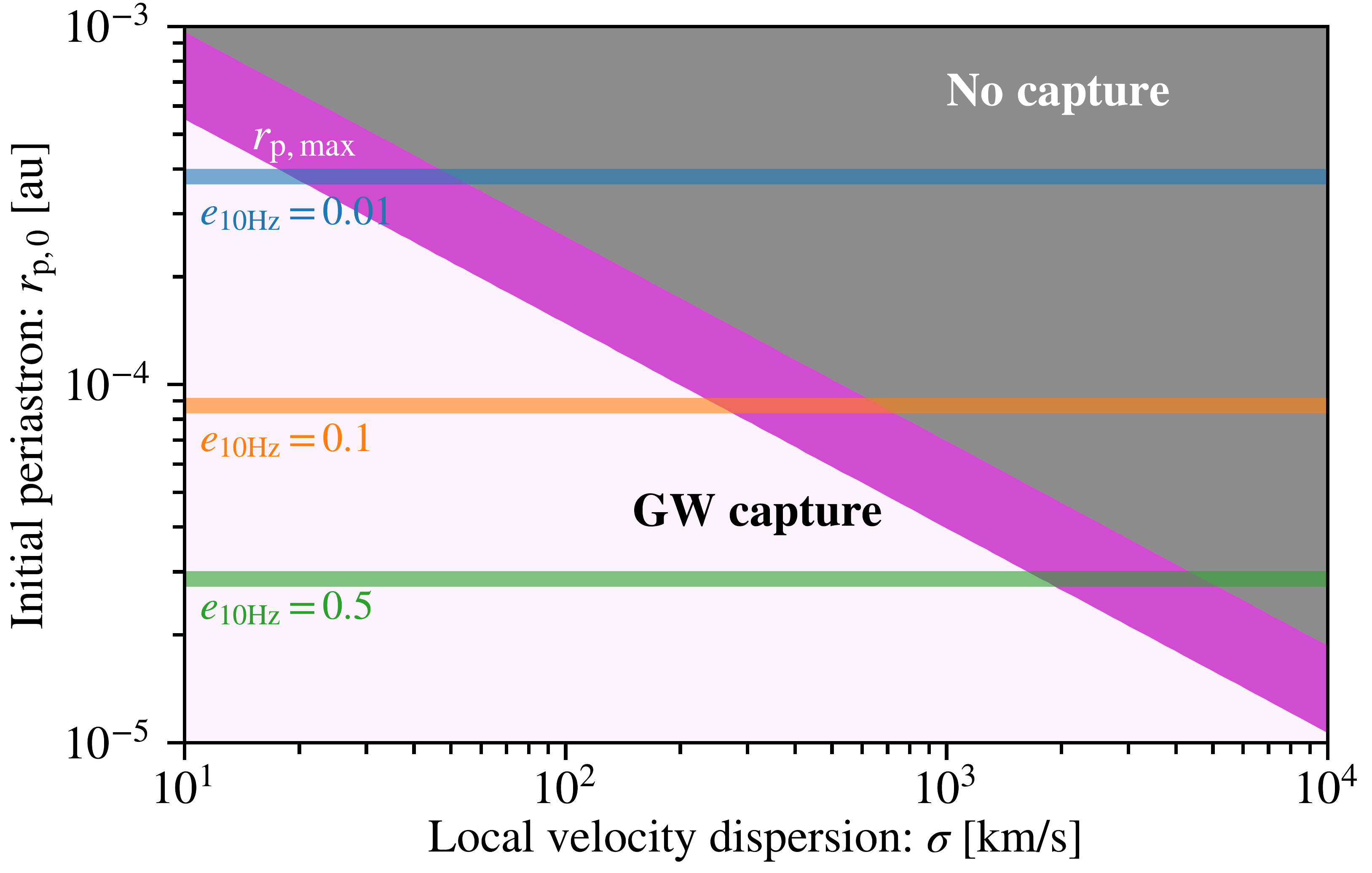} 
\caption{\label{fig:rp_sigma} 
The space of initial periapsis distances and local velocity dispersions for which GW capture is successful (light magenta region) or unsuccessful (grey region).
The dark magenta region corresponds to the 90\% confidence interval of the maximum allowed periastron distance for successful GW capture (\autoref{eq:rp}). 
The blue, orange, and green bands correspond to the 90\% confidence intervals of the initial periastron distances (\autoref{eq:form}) given an eccentricity at the rest-frame GW frequency $f_{\rm GW,r} = 10$ Hz (\autoref{eq:rest}) and using the LVC's mass posteriors GW190521.  
}
\end{figure}

For globular clusters in the Milky Way, the typical one-dimensional velocity dispersions range from $\sim 1$ to $25~{\rm km}/{\rm s}$ \citep{2018MNRAS.478.1520B}, while for nuclear star clusters (without central BHs) the values can range from $\sim 25$ to $35~{\rm km}/{\rm s}$ \citep[e.g.,][]{2005ApJ...618..237W,2008ApJ...687..997S}. For nuclear star clusters with central massive BHs, the velocity dispersion increases closer to the BH \citep[providing a direct relationship between $\sigma$ and binary eccentricity,][]{gondan_eccentric_2018}, while within AGN disks, the velocity dispersion is thought to be some fraction $\sim 0.2$ of the local Keplerian velocity \citep[based on the vector resonant relaxation of BH disks,][]{2018PhRvL.121j1101S,tagawa_eccentric_2020}, meaning the dispersions could range from $\sim10^2$ to $\gtrsim 10^3~{\rm km}/{\rm s}$.  To better understand the space of allowed two-body BH captures, we plot the 90\% confidence interval (dark magenta band) of the maximum allowed periastron distance as a function of the local velocity dispersion in \autoref{fig:rp_sigma} using the LVC mass posteriors for GW190521. 

Given an observed frequency of $f_{\rm GW} = 10 \ {\rm Hz}$ and a lower bound on the eccentricity, we can estimate the corresponding periastron distances at lower GW frequencies.  Assuming an estimated lower bound of $e_{\rm 10Hz} \gtrsim 0.1$, no captures will occur for local velocity dispersions $\sigma \gtrsim 10^3 \ {\rm km}/{\rm s}$, seemingly ruling out a two-body capture in an AGN disk where the velocity dispersion is high \cite[e.g. in the resonant traps near the central BH,][]{ 2019ApJ...878...85S} with eccentricities $e_{\rm 10Hz} \sim 0.1$ 
However, if the very high eccentricities suggested by \cite{gayathri_gw190521_2020} are correct, then two-body captures in any dynamical environment could have formed GW190521.  
%
\section{GW Captures during Three-Body Encounters}
While two-body captures can operate to create BBHs, one of the primary ways to form highly-eccentric mergers from second-generation BHs is during interactions between a BBH and a third BH.  
During these encounters (with velocity dispersions of $\sim 100~{\rm km}/{\rm s}$), the many resonant oscillations of the three bodies offer many opportunities for the close pericenter passages required for GW emission \cite[e.g.,][]{2006ApJ...640..156G,Samsing2014,Samsing2017,Rodriguez2018}.  These encounters can occur in many dynamical environments, such as globular clusters and AGN disks \cite[e.g.,][]{tagawa_eccentric_2020,2020arXiv201009765S}.
\par
To better understand the formation of GW190521-like binaries during GW captures, we focus specifically on formation in globular clusters.  We run a suite of binary-single scatterings using \texttt{fewbody}, a gravitational dynamics integrator for small-$N$ dynamics \citep{Fregeau2007}.  
In addition to Newtonian dynamics, we include the 2.5 post-Newtonian correction to the equations of motion, accounting for GW emission from the system \citep{Antognini2014,Amaro-Seoane2016,Rodriguez2018}.  
This code allows us to track the dynamical properties of BBHs all the way from their dynamical formation to their merger at a distance of $10G(m_1+m_2)/c^2$, where $m_1$ and $m_2$ are the masses of the two components.
\par
The initial conditions for the binary-single scatterings are taken directly from star-by-star models of dense star clusters generated with the Cluster Monte Carlo code, \texttt{CMC} \citep{Joshi2000,Pattabiraman2013}.  We use the suite of models originally developed for \cite{rodriguez_post-newtonian_2018,Rodriguez2019}, which include all the necessary physics for modeling the overall evolution of massive star clusters and their BH and BBH populations, including the aforementioned post-Newtonian corrections.  We identify from those models every binary-single scattering which has at least one component consistent with the $m_1$ and $m_2$ posterior mass distributions for GW190521 at the 90\% confidence level.  
\par
Each encounter is run 100 times with different binary orientations and initial phases (consistent with the implementation in \texttt{CMC}), while the binary separations, eccentricities, velocities, and impact parameters are held fixed.  The eccentricites at a GW frequency of $10$ Hz \cite[consistent with the measurement in][]{romero-shaw_gw190521_2020} are determined by integrating the time-averaged change in semi-major axis and eccentricity \cite[from][]{peters_gravitational_1964} from the point of binary formation until the peak of the rest-frame GW frequency \citep{wen_eccentricity_2003} equals $10 \ {\rm Hz}$. 
See \cite[][Section IID]{rodriguez_post-newtonian_2018} for details.   
\begin{figure}
\centering
\includegraphics[width=\columnwidth]{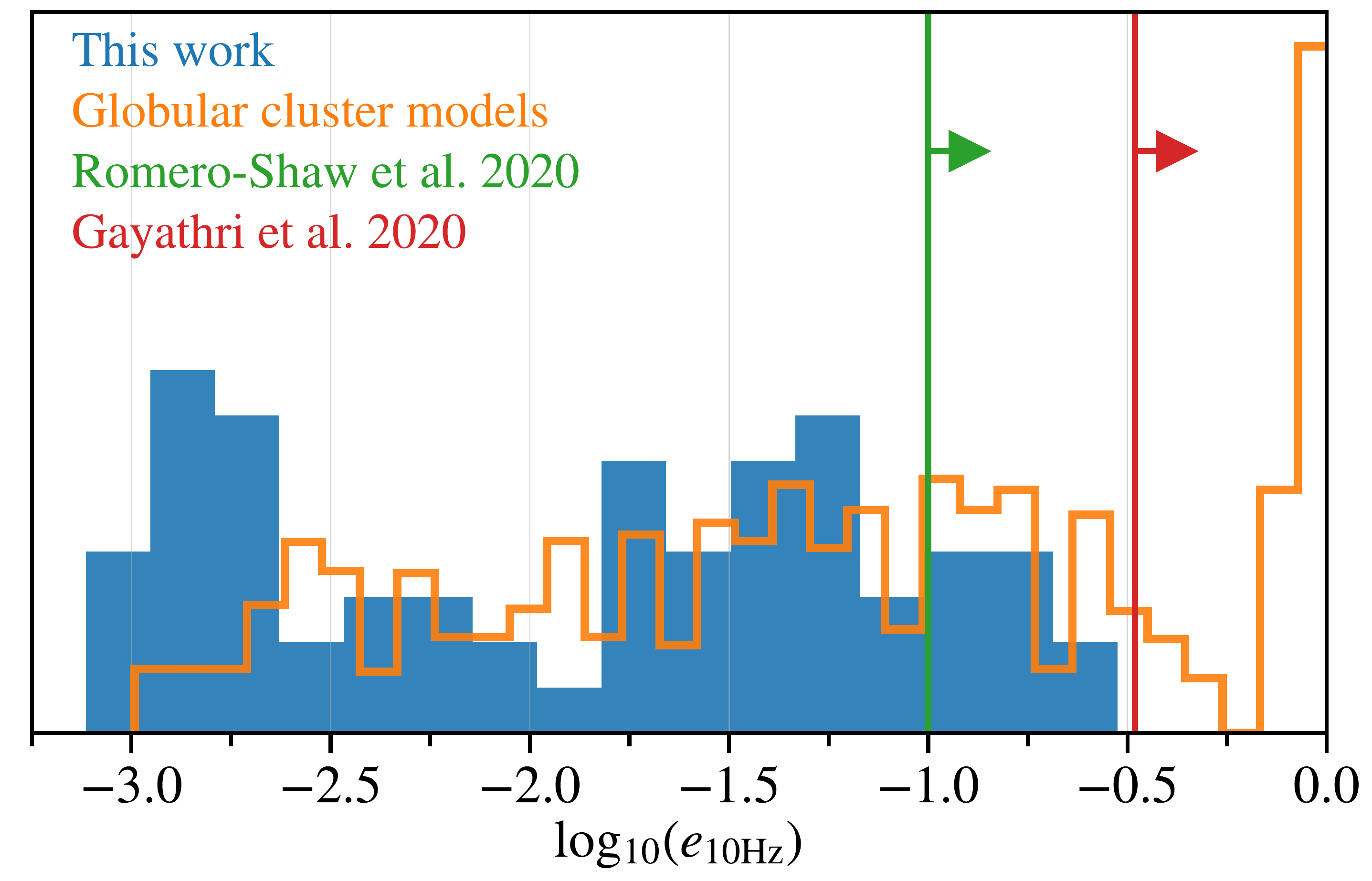} 
\caption{\label{fig:analogues} Distributions of $e_{\rm 10Hz}$, the eccentricity at the rest-frame GW frequency $f_{\rm GW,r} = 10$ Hz. 
The blue distribution corresponds to are our GW190521 analogues obtained from our \texttt{fewbody} binary-single scattering experiments. 
The orange distribution corresponds to stellar-mass BBHs from globular-cluster population models \citep{rodriguez_post-newtonian_2018}. 
The green line corresponds to the lower limit on $e_{\rm 10 Hz}$ that \cite{romero-shaw_gw190521_2020} obtain using the SEOBNRE waveform model.
The red line corresponds to the lower bound on the range of eccentric numerical-relativity waveforms from \cite{gayathri_gw190521_2020} that are consistent with the LVC strain data.
}
\end{figure}
\par
We plot the distribution of the eccentricity at $f_{\rm GW,r} = 10 \ {\rm Hz}$ for our GW190521 analogues in \autoref{fig:analogues} and compare them to the constraints from \cite{romero-shaw_gw190521_2020}, \cite{gayathri_gw190521_2020}, and from the predicted $e_{\rm 10Hz}$ distribution for stellar-mass BBHs from globular cluster population models \citep{rodriguez_post-newtonian_2018}. 
Our analogues have eccentricities at 10Hz that span a broad range, where the majority have effectively circularized, while a smaller subsample have eccentricities $e_{\rm 10Hz}\gtrsim 0.1$. 
Our analogues also demonstrate that binary-single scattering is a viable explanation for GW190521's properties and is consistent with the different studies from the \cite{ligo_scientific_collaboration_and_virgo_collaboration_gw190521_2020}, \cite{romero-shaw_gw190521_2020}.

The eccentricities of our GW190521 analogues, however, do not approach the highly eccentric, i.e., $e \gtrsim 0.9$ regime within the \cite{gayathri_gw190521_2020} constraints, and we find no GW captures where $e_{\rm 10Hz} \gtrsim 0.3$ for any of these systems.  This is in direct contrast to the globular cluster models presented in \cite{rodriguez_post-newtonian_2018}, where a significant fraction of binaries formed with eccentricities of 0.9 or greater (with some binaries \emph{forming} with peak frequencies greater than $10~{\rm Hz}$).  This difference likely arises from the difference in velocity dispersion for the encounters, with more massive BHs having lower typical orbital speeds for the same binding energy (which is what determines the binary's eventual fate in the cluster).  
This suggests that GW190521-like binaries may be less astrophyiscally likely to be highly eccentric $(e \gtrsim 0.9)$ if they form via binary-single scattering. 
\section{Decihertz GW Astronomy}
\subsection{Detectability} 
Even with the LIGO-Virgo network currently operating and with LISA planned for the 2030s, there still exists a frequency gap between these bands of the GW spectrum.
There have been several proposals for a decihertz GW observatory that would bridge the gap between LISA and LIGO-Virgo \citep[e.g.,][]{mandel_astrophysical_2018,canuel_exploring_2018,zhan_zaiga_2019,kuns_astrophysics_2020,kawamura_current_2020,badurina_aion_2020} and contribute to multiband GW observations \citep[e.g.,][]{chen_revealing_2017,ellis_probes_2020}.  
One such ground-based experiment includes the Midband Atomic Gravitational-wave Interferometric Sensor (MAGIS), which uses atom interferometry for GW detection amongst other applications. 
A 100-meter pathfinder experiment is currently being developed \citep{coleman_magis-100_2019}, which will test the technologies necessary for scaling up to a kilometer-sized detector.
\par
We consider here the science achievable for GW190521-like binaries with such a km-scale detector, where our estimates will be more conservative compared to cases where one considers multiple terrestrial detectors or space-based decihertz observatories.   
We consider the projected sensitivity for a km-scale MAGIS detector \citep[e.g.,][]{graham_localizing_2018}. 
\par
With the LVC's constraints on GW190521's parameters, we find that the signal-to-noise ratios (SNRs) for a circular progenitor will be at a sub-threshold level, i.e., ${\rm SNR} \lesssim 5$ for both MAGIS-km and LISA. 
\textit{What about the prospects for GW190521-like binaries?} 
The LVC has provided an event-rate estimate of $0.13_{-0.11}^{+0.30} \ {\rm Gpc}^{-3} \, {\rm yr}^{-1}$ for such sources.
If the progenitors are circular at formation, then LISA could detect ${\sim}1-10$ such events out to $z \lesssim $ over 5-10 years \citep{toubiana_detectable_2020} and similarly for MAGIS-km. 
We show, however, that a dynamical formation for GW190521-like binaries via binary-single scattering may cause them to skip the LISA band or both the LISA and MAGIS bands entirely. 
\par
\begin{figure}
\centering
\includegraphics[width=\columnwidth]{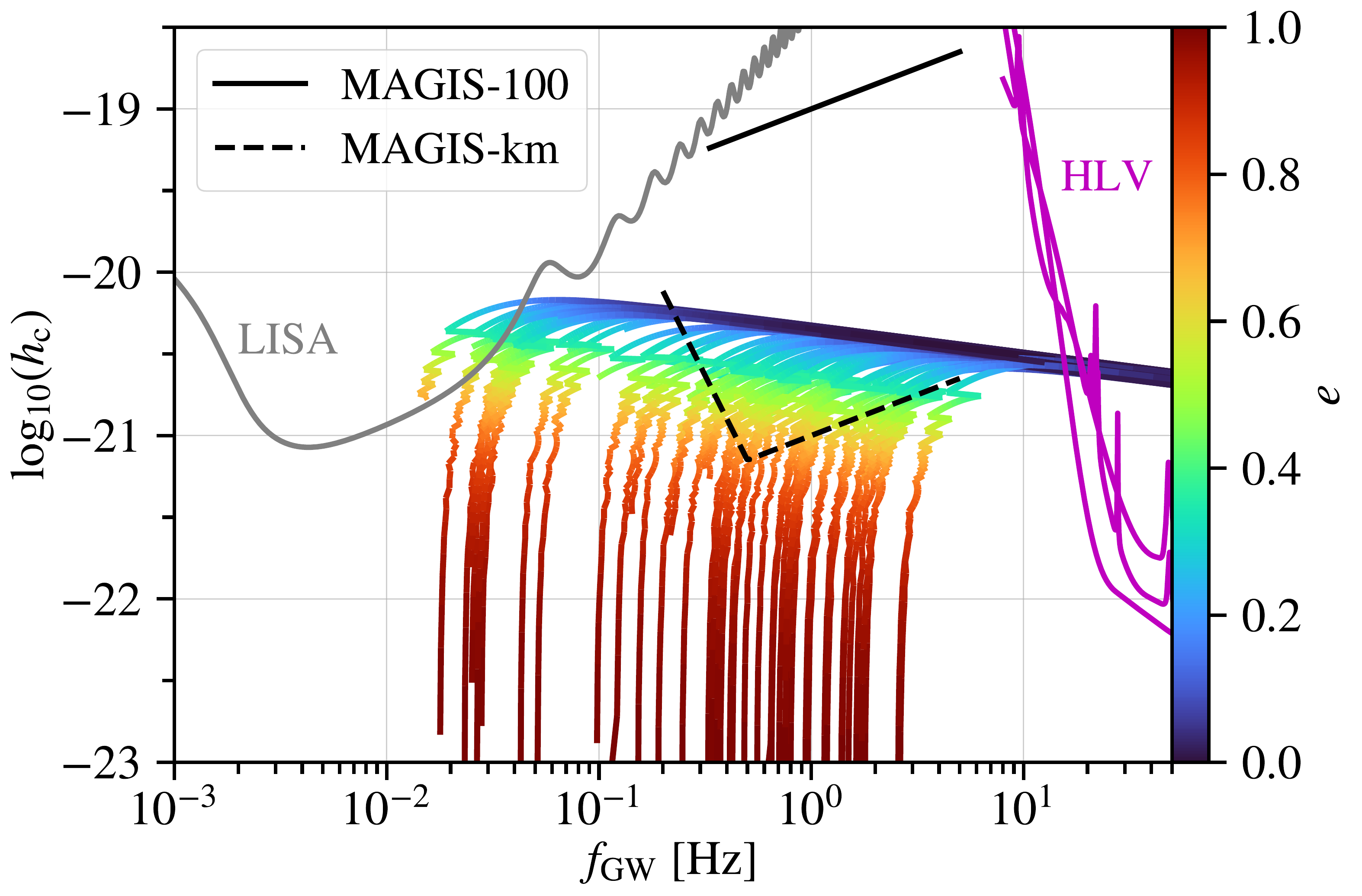}
\includegraphics[width=\columnwidth]{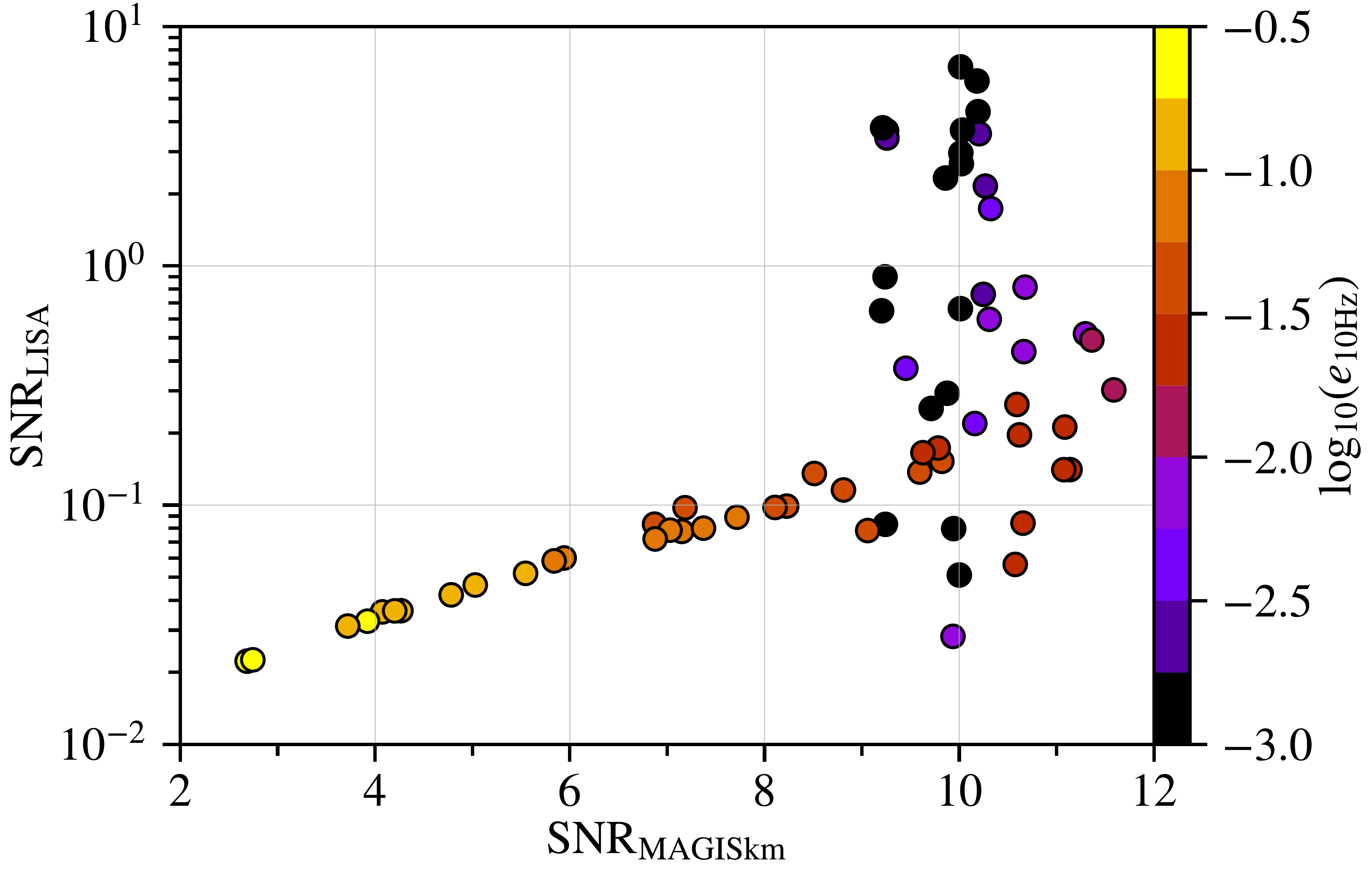}
\caption{\label{fig:hc_snr} Top panel: characteristic strain tracks (\autoref{app:a}) of the peak harmonic ($\max(h_{{\rm c},n}))$ from our \texttt{fewbody} binary-single scattering experiments are plotted as colored lines for a source distance of $D = 1.0 \ {\rm Gpc}$ ($z \approx 0.2$). 
The color corresponds to the eccentricity of that binary as it inspirals towards merger. 
We also plot the detector sensitivities for LISA (gray), MAGIS-100 (solid black), MAGIS-km (dashed black), and the Hanford-Livingston-Virgo (HLV) network (magenta).
Bottom panel: SNR for LISA vs. the SNR in MAGIS-km calculated from the characteristic strain tracks in the top panel and taking the sources to be optimally oriented (\autoref{app:a}). 
The color corresponds to the eccentricity at $f_{\rm GW,r} = 10 \ {\rm Hz}$ for that sample. 
}
\end{figure}

We plot the characteristic-strain tracks (details in \autoref{app:a}) of the peak harmonic with the sensitivities of aLIGO, MAGIS-100, MAGIS-km, and LISA in the top panel of \autoref{fig:hc_snr}, assuming an optimal source orientation. 
Our GW190521 analogues form over a wide range of $f_{\rm GW}$, where they can form in the LISA band, the MAGIS-km band, or skip both bands entirely. 
We plot in the bottom panel of \autoref{fig:hc_snr} the SNRs for LISA and for MAGIS-km. 
From our \texttt{fewbody} binary-single scattering events, we find no GW190521 analogues that are detectable in the LISA band  (SNR$_{\rm LISA} < 8$) for source distances of $D = 1.0 \ {\rm Gpc}$ (corresponding to $z \approx 0.2$).
We do, however, find two samples that range within $5 < {\rm SNR}_{\rm LISA} < 8$, which may be of interest for multiband GW follow-up analysis with MAGIS-km and LIGO-Virgo. 
The SNRs in the MAGIS band that are $> 8$ have eccentricities at $f_{\rm GW} = 10 \, {\rm Hz}$ that are $e_{\rm 10Hz} < 0.1$, lower than both the \cite{romero-shaw_gw190521_2020} and \cite{gayathri_gw190521_2020} constraints. 
\subsection{Eccentricity constraints}
\begin{figure}
\centering
\includegraphics[width=\columnwidth]{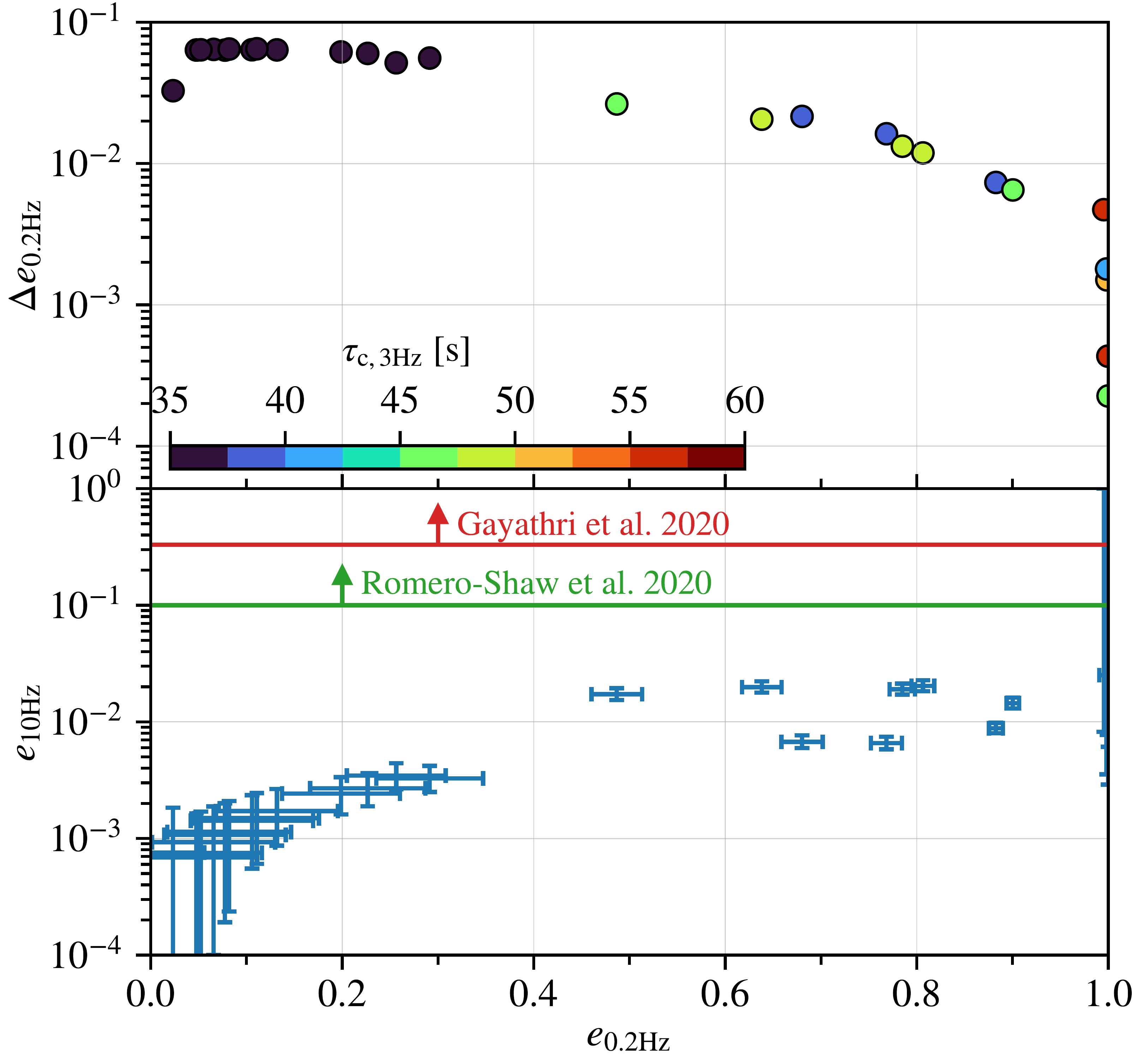} 
\caption{\label{fig:delta_e10} Top panel: Fisher estimates (\autoref{app:a}) of $\Delta e_{\rm 0.2Hz}$, the uncertainty in the eccentricity at $f_{\rm GW,r} = 0.2$, vs.~$e_{\rm 0.2Hz}$ from our GW190521 analogues with SNR $\ge 10$. 
The color of each sample corresponds to the coalescence timescale at 3 Hz \citep{peters_gravitational_1964,zwick_improved_2020}. 
Bottom panel: The constraints on $e_{\rm 10Hz}$ obtained by propagating $e_{\rm 0.2Hz}$ and its Fisher uncertainties via \cite{peters_gravitational_1963}. 
Our highly eccentric samples, $e_{\rm 0.2Hz} > 0.99$, have Fisher uncertainties that make $e_{\rm 0.2Hz} = 1$ as the upper bound, such that $e_{\rm 10Hz}$ may be highly eccentric as well. 
The lower limits on $e_{\rm 10Hz}$ for GW190521 from \cite{romero-shaw_gw190521_2020} and \cite{gayathri_gw190521_2020} are shown as the green and red lines, respectively. 
}
\end{figure}

To estimate how much the constraints on the binary eccentricity can be improved, we carry out a Fisher-information-matrix analysis using our samples that have SNR $\ge 10$ in the MAGIS-km detector. 
In the top panel of \autoref{fig:delta_e10}, we plot the Fisher estimates of the eccentricity uncertainties $\Delta e_{\rm 0.2Hz}$ as the binary enters the MAGIS band at $f_{\rm GW} = 0.2$ Hz for samples that have ${\rm SNR} \ge 10$. 
%
From the quadrupole approximation, the coalescence timescale for an eccentric binary given a reference $a$ and $e$ is \citep{peters_gravitational_1964}
\begin{equation}
\tau_{\rm c} = \frac{12}{19} \frac{\alpha^4}{\beta} \int_0^{e_0} \frac{e^{29/19} \left[1 + \frac{121}{304} e^2\right]^{1181/2299}}{\left(1 - e^2\right)^{3/2}} \ {\rm d}e\ .
\end{equation}
where the quantities $\alpha$ and $\beta$ are defined as 
\begin{subequations}
\begin{align}
\alpha &= \frac{a_0 \left(1 - e_0^2\right)}{e_0^{12/19}} \left(1 + \frac{121}{304}e_0^2\right)^{-870/2299} \ , \\
\beta &= \frac{64}{5}\frac{G^3}{c^5} m_1 m_2 \left(m_1 + m_2\right) \ .
\end{align}
\end{subequations}
\cite{zwick_improved_2020} have provided correction factors to improve this estimate of the coalescence timescale, which we incorporate into our calculations and plot as the color of each sample in the top panel of \autoref{fig:delta_e10}. 
The coalescence timescales of these detectable GW190521 analogues occur on sub-minute timescales, such that the sky-area localization would not be well-constrained with just a single ground-based decihertz GW detector.
\par
The Fisher estimates at low eccentricities are such that the observed GW signal would be consistent with $e_{\rm 0.2Hz} = 0$, while the uncertainty $\Delta e_{\rm 0.2Hz}$ decreases at higher eccentricities. 
The constraints on $e_{\rm 0.2Hz}$ can then be propagated to higher frequencies via \cite{peters_gravitational_1963} to obtain constraints on $e_{\rm 10Hz}$, which we plot in the bottom panel of \autoref{fig:delta_e10}. 
Here, the $x$-axis errorbars are the Fisher estimates from the top panel and the $y$-axis errorbars are obtained from the error propagation. 
Our highly eccentric samples, $e_{\rm 0.2Hz} > 0.99$, have Fisher uncertainties that make $e_{\rm 0.2Hz} = 1$ the upper bound, such that the $e_{\rm 10Hz}$ may be highly eccentric as well. 
With LIGO-Virgo observations alone, the $e_{\rm 10Hz} > 0$ scenario cannot necessarily be favored over $e_{\rm 10Hz} = 0$ with spin precession.
Pre-merger MAGIS-km observations would be able to distinguish between these two scenarios and joint MAGIS+LIGO-Virgo observations can be used for multiband GW parameter estimation. 
\section{Discussion and conclusions} \label{sec:conclusions}
GW190521's detection provides novel constraints on astrophysical formation channels. 
With independent claims of finite eccentricity for GW190521 as it entered the LIGO band, we investigate the implications for the dynamical formation scenarios of GW single-single capture and binary-single scattering. 
For GW capture, we constrain the parameter space of initial periapses and local velocity dispersions that can produce successful captures. 
Such constraints can then be mapped to global models of globular clusters and nuclear star clusters. 
If GW190521 had $e_{\rm 10Hz} \gtrsim 0.1$, it would have been unlikely to form via GW capture in high velocity-dispersion environments with $\sigma \gtrsim 10^3 \ {\rm km} \, {\rm s}^{-1}$, i.e., within the broad-line region of AGN.
The AGN-disk channel, however, may still be a viable formation scenario for GW190521-like binaries. 
\par
We instead consider a  binary-single scattering origin and model the dynamical formation of GW190521-like binaries with the \texttt{fewbody} code. 
The majority of our binaries have effectively circularized as they reach $f_{\rm GW,r} = 10$ Hz, while a smaller subsample have $e_{\rm 10Hz} > 0.1$,  consistent with the reported constraints from \cite{romero-shaw_gw190521_2020}. 
This subsample, however has eccentricities well below what \cite{gayathri_gw190521_2020} suggest in their analysis.
While LIGO-Virgo data itself is insufficient to unambiguously favor $e_{\rm 10Hz} > 0$ over $e_{\rm 10Hz} = 0$, this event further motivates the development of decihertz GW astronomy. 
We find that while LISA may not be sensitive to GW190521-like binaries that form with large eccentricities, a decihertz GW observatory may be able to detect such dynamically formed binaries and provide independent constraints on the eccentricity.
\par
Combining multiband MAGIS+LIGO-Virgo observations of eccentric GW190521-like binaries can provide joint constraints on the eccentricity evolution from formation all the way to merger. 
These joint constraints can provide more informed insights on the possible dynamical formation scenarios that we have discussed here and the viability of alternative formation scenarios, including the isolated binary channel and the AGN disk channel for GW190521-like binaries.
Our models further demonstrate the need for a decihertz GW observatory at the level of MAGIS-km or better in order in to make such science possible. 
Even in the case of a non-detection, a MAGIS-km detector could be able to set an independent lower limit on $e_{\rm 10Hz}$.
\par
Early detection in MAGIS would provide alerts for multi-messenger follow-up to search for a possible electromagnetic counterpart. 
The sky-area localization, however, would not be well-constrained with a single baseline since the merger occurs on sub-minute timescales. 
This would motivate a global network of 2-3 ground-based atom-interferometric detectors in order localize at a similar precision as LIGO for decihertz GW signals that occur on sub-minute timescales. 
%
\acknowledgments
AMH is supported by the McWilliams Postdoctoral Fellowship. 
We thank Pau Amaro-Seoane for helpful comments that improved this manuscript. 
This work used the LVC's publicly available GW190521 posteriors. 


%



 \software{\texttt{fewbody} \citep{Fregeau2007}, \texttt{numpy} \citep{walt_numpy_2011}, \texttt{scipy} \citep{2020SciPy}, \texttt{matplotlib} \citep{hunter_matplotlib:_2007}}.

\bibliographystyle{yahapj}
\bibliography{references}

\begin{thebibliography}{}
\providecommand\natexlab[1]{#1}
\providecommand\JournalTitle[1]{#1}

\bibitem[{{Abbott} {et~al.}(2020){Abbott}, {Abbott}, {Abraham}, {Acernese},
  {Ackley}, {Adams}, {Adhikari}, {Adya}, {Affeldt}, {Agathos}, {Agatsuma},
  {Aggarwal}, {Aguiar}, {Aich}, {Aiello}, {Jadhav}, {James}, {Jani},
  {Janthalur}, {Jaranowski}, {Jariwala}, {Jaume}, {Jenkins}, {Jiang}, {LIGO
  Scientific Collaboration}, \& {Virgo Collaboration}}]{2020ApJ...900L..13A}
{Abbott}, R., {Abbott}, T.~D., {Abraham}, S., {et~al.} 2020,
  \href{http://dx.doi.org/10.3847/2041-8213/aba493}{\JournalTitle{\apjl}, 900,
  L13}

\bibitem[{{Amaro-Seoane} \& Chen(2016)}]{Amaro-Seoane2016}
{Amaro-Seoane}, P., \& Chen, X. 2016,
  \href{http://dx.doi.org/10.1093/mnras/stw503}{\JournalTitle{MNRAS}, 458,
  3075}

\bibitem[{Amaro-Seoane \& Santamar\'{i}a(2010)}]{amaro-seoane_detection_2010}
Amaro-Seoane, P., \& Santamar\'{i}a, L. 2010,
  \href{http://dx.doi.org/10.1088/0004-637X/722/2/1197}{\JournalTitle{ApJ},
  722, 1197}

\bibitem[{Amaro-Seoane {et~al.}(2017)Amaro-Seoane, Audley, Babak, Baker,
  Barausse, Bender, Berti, Binetruy, Born, Bortoluzzi, Camp, Caprini, Cardoso,
  Colpi, Conklin, Cornish, Cutler, Danzmann, Dolesi, Ferraioli, Ferroni,
  Fitzsimons, Gair, Bote, Giardini, Gibert, Grimani, Halloin, Heinzel, Hertog,
  Hewitson, Holley-Bockelmann, Hollington, Hueller, Inchauspe, Jetzer,
  Karnesis, Killow, Klein, Klipstein, Korsakova, Larson, Livas, Lloro, Man,
  Mance, Martino, Mateos, McKenzie, McWilliams, Miller, Mueller, Nardini,
  Nelemans, Nofrarias, Petiteau, Pivato, Plagnol, Porter, Reiche, Robertson,
  Robertson, Rossi, Russano, Schutz, Sesana, Shoemaker, Slutsky, Sopuerta,
  Sumner, Tamanini, Thorpe, Troebs, Vallisneri, Vecchio, Vetrugno, Vitale,
  Volonteri, Wanner, Ward, Wass, Weber, Ziemer, \&
  Zweifel}]{amaro-seoane_laser_2017}
Amaro-Seoane, P., Audley, H., Babak, S., {et~al.} 2017,
  \href{http://arxiv.org/abs/1702.00786}{\JournalTitle{arXiv:1702.00786
  [astro-ph]}}, arXiv: 1702.00786

\bibitem[{Antognini {et~al.}(2014)Antognini, Shappee, Thompson, \&
  {Amaro-Seoane}}]{Antognini2014}
Antognini, J.~M., Shappee, B.~J., Thompson, T.~A., \& {Amaro-Seoane}, P. 2014,
  \href{http://dx.doi.org/10.1093/mnras/stu039}{\JournalTitle{MNRAS}, 439,
  1079}

\bibitem[{{Antonini} {et~al.}(2014){Antonini}, {Murray}, \&
  {Mikkola}}]{2014ApJ...781...45A}
{Antonini}, F., {Murray}, N., \& {Mikkola}, S. 2014,
  \href{http://dx.doi.org/10.1088/0004-637X/781/1/45}{\JournalTitle{\apj}, 781,
  45}

\bibitem[{{Antonini} \& {Perets}(2012)}]{2012ApJ...757...27A}
{Antonini}, F., \& {Perets}, H.~B. 2012,
  \href{http://dx.doi.org/10.1088/0004-637X/757/1/27}{\JournalTitle{\apj}, 757,
  27}

\bibitem[{{Arca-Sedda} {et~al.}(2018){Arca-Sedda}, {Li}, \&
  {Kocsis}}]{2018arXiv180506458A}
{Arca-Sedda}, M., {Li}, G., \& {Kocsis}, B. 2018, \JournalTitle{arXiv
  e-prints}, arXiv:1805.06458

\bibitem[{Badurina {et~al.}(2020)Badurina, Bentine, Blas, Bongs, Bortoletto,
  Bowcock, Bridges, Bowden, Buchmueller, Burrage, Coleman, Elertas, Ellis,
  Foot, Gibson, Haehnelt, Harte, Hedges, Hobson, Holynski, Jones, Langlois,
  Lellouch, Lewicki, Maiolino, Majewski, Malik, March-Russell, McCabe, Newbold,
  Sauer, Schneider, Shipsey, Singh, Uchida, Valenzuela, Grinten, Vaskonen,
  Vossebeld, Weatherill, \& Wilmut}]{badurina_aion_2020}
Badurina, L., Bentine, E., Blas, D., {et~al.} 2020,
  \href{http://dx.doi.org/10.1088/1475-7516/2020/05/011}{\JournalTitle{J.
  Cosmol. Astropart. Phys.}, 2020, 011}

\bibitem[{Barausse {et~al.}(2014)Barausse, Cardoso, \&
  Pani}]{barausse_can_2014}
Barausse, E., Cardoso, V., \& Pani, P. 2014,
  \href{http://dx.doi.org/10.1103/PhysRevD.89.104059}{\JournalTitle{PRD}, 89,
  104059}

\bibitem[{{Baumgardt} \& {Hilker}(2018)}]{2018MNRAS.478.1520B}
{Baumgardt}, H., \& {Hilker}, M. 2018,
  \href{http://dx.doi.org/10.1093/mnras/sty1057}{\JournalTitle{\mnras}, 478,
  1520}

\bibitem[{Belczynski(2020)}]{belczynski_most_2020}
Belczynski, K. 2020,
  \href{http://arxiv.org/abs/2009.13526}{\JournalTitle{arXiv:2009.13526
  [astro-ph]}}, arXiv: 2009.13526

\bibitem[{Berry \& Gair(2010)}]{berry_gravitational_2010}
Berry, C. P.~L., \& Gair, J.~R. 2010,
  \href{http://dx.doi.org/10.1103/PhysRevD.82.107501}{\JournalTitle{Phys. Rev.
  D}, 82, 107501}

\bibitem[{Canuel {et~al.}(2018)Canuel, Bertoldi, Amand, Pozzo~di Borgo,
  Chantrait, Danquigny, Dovale~Álvarez, Fang, Freise, Geiger, Gillot, Henry,
  Hinderer, Holleville, Junca, Lefèvre, Merzougui, Mielec, Monfret, Pelisson,
  Prevedelli, Reynaud, Riou, Rogister, Rosat, Cormier, Landragin, Chaibi,
  Gaffet, \& Bouyer}]{canuel_exploring_2018}
Canuel, B., Bertoldi, A., Amand, L., {et~al.} 2018,
  \href{http://dx.doi.org/10.1038/s41598-018-32165-z}{\JournalTitle{Scientific
  Reports}, 8, 14064}

\bibitem[{Caputo {et~al.}(2020)Caputo, Sberna, Toubiana, Babak, Barausse,
  Marsat, \& Pani}]{caputo_gravitational-wave_2020}
Caputo, A., Sberna, L., Toubiana, A., {et~al.} 2020,
  \href{http://dx.doi.org/10.3847/1538-4357/ab7b66}{\JournalTitle{ApJ}, 892,
  90}

\bibitem[{Chen \& Amaro-Seoane(2017)}]{chen_revealing_2017}
Chen, X., \& Amaro-Seoane, P. 2017,
  \href{http://dx.doi.org/10.3847/2041-8213/aa74ce}{\JournalTitle{ApJL}, 842,
  L2}

\bibitem[{Coleman(2019)}]{coleman_magis-100_2019}
Coleman, J. 2019, \href{http://dx.doi.org/10.22323/1.340.0021}{in Proceedings
  of {The} 39th {International} {Conference} on {High} {Energy} {Physics} —
  {PoS}({ICHEP2018}), Vol. 340} (SISSA Medialab), 021

\bibitem[{Costa {et~al.}(2020)Costa, Bressan, Mapelli, Marigo, Iorio, \&
  Spera}]{costa_formation_2020}
Costa, G., Bressan, A., Mapelli, M., {et~al.} 2020,
  \href{http://adsabs.harvard.edu/abs/2020arXiv201002242C}{\JournalTitle{arXiv
  e-prints}, 2010, arXiv:2010.02242}

\bibitem[{D'Orazio \& Samsing(2018)}]{dorazio_black_2018}
D'Orazio, D.~J., \& Samsing, J. 2018,
  \href{http://dx.doi.org/10.1093/mnras/sty2568}{\JournalTitle{MNRAS}, 481,
  4775}

\bibitem[{Ellis \& Vaskonen(2020)}]{ellis_probes_2020}
Ellis, J., \& Vaskonen, V. 2020,
  \href{http://dx.doi.org/10.1103/PhysRevD.101.124013}{\JournalTitle{Phys. Rev.
  D}, 101, 124013}

\bibitem[{Finn(1996)}]{Finn1996}
Finn, L.~S. 1996,
  \href{http://dx.doi.org/10.1103/PhysRevD.53.2878}{\JournalTitle{Physical
  Review D}, 53, 2878}

\bibitem[{{Fragione} {et~al.}(2020){Fragione}, {Loeb}, \&
  {Rasio}}]{2020ApJ...902L..26F}
{Fragione}, G., {Loeb}, A., \& {Rasio}, F.~A. 2020,
  \href{http://dx.doi.org/10.3847/2041-8213/abbc0a}{\JournalTitle{\apjl}, 902,
  L26}

\bibitem[{Fregeau \& Rasio(2007)}]{Fregeau2007}
Fregeau, J.~M., \& Rasio, F.~A. 2007, \JournalTitle{ApJ}, 658, 1047

\bibitem[{Gayathri {et~al.}(2020)Gayathri, Healy, Lange, O'Brien, Szczepanczyk,
  Bartos, Campanelli, Klimenko, Lousto, \&
  O'Shaughnessy}]{gayathri_gw190521_2020}
Gayathri, V., Healy, J., Lange, J., {et~al.} 2020,
  \href{http://arxiv.org/abs/2009.05461}{\JournalTitle{arXiv:2009.05461
  [astro-ph, physics:gr-qc]}}, arXiv: 2009.05461

\bibitem[{{Gond{\'a}n} \& {Kocsis}(2020)}]{2020arXiv201102507G}
{Gond{\'a}n}, L., \& {Kocsis}, B. 2020, \JournalTitle{arXiv e-prints},
  arXiv:2011.02507

\bibitem[{Gond\'{a}n {et~al.}(2018)Gond\'{a}n, Kocsis, Raffai, \&
  Frei}]{gondan_eccentric_2018}
Gond\'{a}n, L., Kocsis, B., Raffai, P., \& Frei, Z. 2018,
  \href{http://dx.doi.org/10.3847/1538-4357/aabfee}{\JournalTitle{ApJ}, 860, 5}

\bibitem[{Graham \& Jung(2018)}]{graham_localizing_2018}
Graham, P.~W., \& Jung, S. 2018,
  \href{http://dx.doi.org/10.1103/PhysRevD.97.024052}{\JournalTitle{Phys. Rev.
  D}, 97, 024052}

\bibitem[{{G{\"u}ltekin} {et~al.}(2006){G{\"u}ltekin}, {Miller}, \&
  {Hamilton}}]{2006ApJ...640..156G}
{G{\"u}ltekin}, K., {Miller}, M.~C., \& {Hamilton}, D.~P. 2006,
  \href{http://dx.doi.org/10.1086/499917}{\JournalTitle{\apj}, 640, 156}

\bibitem[{{Hoang} {et~al.}(2018){Hoang}, {Naoz}, {Kocsis}, {Rasio}, \&
  {Dosopoulou}}]{2018ApJ...856..140H}
{Hoang}, B.-M., {Naoz}, S., {Kocsis}, B., {Rasio}, F.~A., \& {Dosopoulou}, F.
  2018, \href{http://dx.doi.org/10.3847/1538-4357/aaafce}{\JournalTitle{\apj},
  856, 140}

\bibitem[{Holgado \& Ricker(2019)}]{holgado_gravitational_2019}
Holgado, A.~M., \& Ricker, P.~M. 2019,
  \href{http://dx.doi.org/10.3847/1538-4357/ab3293}{\JournalTitle{ApJ}, 882,
  39}

\bibitem[{Huerta {et~al.}(2015)Huerta, McWilliams, Gair, \&
  Taylor}]{huerta_detection_2015}
Huerta, E., McWilliams, S.~T., Gair, J.~R., \& Taylor, S.~R. 2015,
  \href{http://dx.doi.org/10.1103/PhysRevD.92.063010}{\JournalTitle{Phys. Rev.
  D}, 92, 063010}

\bibitem[{Hunter(2007)}]{hunter_matplotlib:_2007}
Hunter, J.~D. 2007,
  \href{http://dx.doi.org/10.1109/MCSE.2007.55}{\JournalTitle{Computing in
  Science Engineering}, 9, 90}

\bibitem[{Joshi {et~al.}(2000)Joshi, Rasio, Zwart, \&
  Portegies~Zwart}]{Joshi2000}
Joshi, K.~J., Rasio, F.~A., Zwart, S.~P., \& Portegies~Zwart, S. 2000,
  \href{http://dx.doi.org/10.1086/309350}{\JournalTitle{ApJ}, 540, 969}

\bibitem[{Kawamura {et~al.}(2020)Kawamura, Ando, Seto, Sato, Musha, Kawano,
  Yokoyama, Tanaka, Ioka, Akutsu, Takashima, Agatsuma, Araya, Aritomi, Asada,
  Chiba, Eguchi, Enoki, Fujimoto, Fujita, Futamase, Harada, Hayama, Himemoto,
  Hiramatsu, Hong, Hosokawa, Ichiki, Ikari, Ishihara, Ishikawa, Itoh, Ito,
  Iwaguchi, Izumi, Kanda, Kanemura, Kawazoe, Kobayashi, Kohri, Kojima,
  Kokeyama, Kotake, Kuroyanagi, Maeda, Matsushita, Michimura, Morimoto,
  Mukohyama, Nagano, Nagano, Naito, Nakamura, Nakamura, Nakano, Nakao,
  Nakasuka, Nakayama, Nakazawa, Nishizawa, Ohkawa, Oohara, Sago, Saijo,
  Sakagami, Sakai, Sato, Shibata, Shinkai, Shoda, Somiya, Sotani, Takahashi,
  Takahashi, Akiteru, Taniguchi, Taruya, Tsubono, Tsujikawa, Ueda, Ueda,
  Watanabe, Yagi, Yamada, Yokoyama, Yoo, \& Zhu}]{kawamura_current_2020}
Kawamura, S., Ando, M., Seto, N., {et~al.} 2020,
  \href{http://arxiv.org/abs/2006.13545}{\JournalTitle{arXiv:2006.13545
  [gr-qc]}}, arXiv: 2006.13545

\bibitem[{{Kimball} {et~al.}(2020){Kimball}, {Talbot}, {Berry}, {Zevin},
  {Thrane}, {Kalogera}, {Buscicchio}, {Carney}, {Dent}, {Middleton}, {Payne},
  {Veitch}, \& {Williams}}]{2020arXiv201105332K}
{Kimball}, C., {Talbot}, C., {Berry}, C. P.~L., {et~al.} 2020,
  \JournalTitle{arXiv e-prints}, arXiv:2011.05332

\bibitem[{{Kremer} {et~al.}(2020){Kremer}, {Spera}, {Becker}, {Chatterjee}, {Di
  Carlo}, {Fragione}, {Rodriguez}, {Ye}, \& {Rasio}}]{2020ApJ...903...45K}
{Kremer}, K., {Spera}, M., {Becker}, D., {et~al.} 2020,
  \href{http://dx.doi.org/10.3847/1538-4357/abb945}{\JournalTitle{\apj}, 903,
  45}

\bibitem[{Kuns {et~al.}(2020)Kuns, Yu, Chen, \&
  Adhikari}]{kuns_astrophysics_2020}
Kuns, K.~A., Yu, H., Chen, Y., \& Adhikari, R.~X. 2020,
  \href{http://dx.doi.org/10.1103/PhysRevD.102.043001}{\JournalTitle{Phys. Rev.
  D}, 102, 043001}

\bibitem[{{Liu} \& {Lai}(2020)}]{2020arXiv200910068L}
{Liu}, B., \& {Lai}, D. 2020, \JournalTitle{arXiv e-prints}, arXiv:2009.10068

\bibitem[{{LVC}(2020{\natexlab{a}})}]{ligo_scientific_collaboration_and_virgo_collaboration_gw190521_2020}
{LVC}. 2020{\natexlab{a}},
  \href{http://dx.doi.org/10.1103/PhysRevLett.125.101102}{\JournalTitle{Phys.
  Rev. Lett.}, 125, 101102}

\bibitem[{{LVC}(2020{\natexlab{b}})}]{abbott_properties_2020}
---. 2020{\natexlab{b}},
  \href{http://dx.doi.org/10.3847/2041-8213/aba493}{\JournalTitle{ApJL}, 900,
  L13}

\bibitem[{Mandel {et~al.}(2018)Mandel, Sesana, \&
  Vecchio}]{mandel_astrophysical_2018}
Mandel, I., Sesana, A., \& Vecchio, A. 2018,
  \href{http://dx.doi.org/10.1088/1361-6382/aaa7e0}{\JournalTitle{Class.
  Quantum Grav.}, 35, 054004}

\bibitem[{Michaely \& Perets(2020)}]{michaely_high_2020}
Michaely, E., \& Perets, H.~B. 2020,
  \href{http://dx.doi.org/10.1093/mnras/staa2720}{\JournalTitle{MNRAS}, 498,
  4924}

\bibitem[{Pattabiraman {et~al.}(2013)Pattabiraman, Umbreit, Liao, Choudhary,
  Kalogera, Memik, \& Rasio}]{Pattabiraman2013}
Pattabiraman, B., Umbreit, S., Liao, W.-k., {et~al.} 2013,
  \href{http://dx.doi.org/10.1088/0067-0049/204/2/15}{\JournalTitle{ApJS}, 204,
  15}

\bibitem[{Peters(1964)}]{peters_gravitational_1964}
Peters, P.~C. 1964,
  \href{http://dx.doi.org/10.1103/PhysRev.136.B1224}{\JournalTitle{Phys. Rev.},
  136, B1224}

\bibitem[{Peters \& Mathews(1963)}]{peters_gravitational_1963}
Peters, P.~C., \& Mathews, J. 1963,
  \href{http://dx.doi.org/10.1103/PhysRev.131.435}{\JournalTitle{Phys. Rev.},
  131, 435}

\bibitem[{Quinlan \& Shapiro(1987)}]{quinlan_collapse_1987}
Quinlan, G.~D., \& Shapiro, S.~L. 1987,
  \href{http://dx.doi.org/10.1086/165624}{\JournalTitle{ApJ}, 321, 199}

\bibitem[{Renzo {et~al.}(2020)Renzo, Cantiello, Metzger, \&
  Jiang}]{renzo_stellar_2020}
Renzo, M., Cantiello, M., Metzger, B.~D., \& Jiang, Y.-F. 2020,
  \href{http://adsabs.harvard.edu/abs/2020arXiv201000705R}{\JournalTitle{arXiv
  e-prints}, 2010, arXiv:2010.00705}

\bibitem[{{Renzo} {et~al.}(2020){Renzo}, {Cantiello}, {Metzger}, \&
  {Jiang}}]{2020ApJ...904L..13R}
{Renzo}, M., {Cantiello}, M., {Metzger}, B.~D., \& {Jiang}, Y.~F. 2020,
  \href{http://dx.doi.org/10.3847/2041-8213/abc6a6}{\JournalTitle{\apjl}, 904,
  L13}

\bibitem[{Rice \& Zhang(2020)}]{rice_growth_2020}
Rice, J.~R., \& Zhang, B. 2020,
  \href{http://adsabs.harvard.edu/abs/2020arXiv200911326R}{\JournalTitle{arXiv
  e-prints}, 2009, arXiv:2009.11326}

\bibitem[{Rodriguez {et~al.}(2018{\natexlab{a}})Rodriguez, Amaro-Seoane,
  Chatterjee, Kremer, Rasio, Samsing, Ye, \&
  Zevin}]{rodriguez_post-newtonian_2018}
Rodriguez, C.~L., Amaro-Seoane, P., Chatterjee, S., {et~al.}
  2018{\natexlab{a}},
  \href{http://dx.doi.org/10.1103/PhysRevD.98.123005}{\JournalTitle{Phys. Rev.
  D}, 98, 123005}

\bibitem[{Rodriguez {et~al.}(2018{\natexlab{b}})Rodriguez, {Amaro-Seoane},
  Chatterjee, \& Rasio}]{Rodriguez2018}
Rodriguez, C.~L., {Amaro-Seoane}, P., Chatterjee, S., \& Rasio, F.~A.
  2018{\natexlab{b}},
  \href{http://dx.doi.org/10.1103/PhysRevLett.120.151101}{\JournalTitle{Phys.
  Rev. Lett.}, 120, 151101}

\bibitem[{Rodriguez {et~al.}(2016)Rodriguez, Chatterjee, \&
  Rasio}]{rodriguez_binary_2016}
Rodriguez, C.~L., Chatterjee, S., \& Rasio, F.~A. 2016,
  \href{http://dx.doi.org/10.1103/PhysRevD.93.084029}{\JournalTitle{Phys. Rev.
  D}, 93, 084029}

\bibitem[{Rodriguez {et~al.}(2015)Rodriguez, Morscher, Pattabiraman,
  Chatterjee, Haster, \& Rasio}]{rodriguez_binary_2015}
Rodriguez, C.~L., Morscher, M., Pattabiraman, B., {et~al.} 2015,
  \href{http://dx.doi.org/10.1103/PhysRevLett.115.051101}{\JournalTitle{Phys.
  Rev. Lett.}, 115, 051101}

\bibitem[{Rodriguez {et~al.}(2019)Rodriguez, Zevin, {Amaro-Seoane}, Chatterjee,
  Kremer, Rasio, \& Ye}]{Rodriguez2019}
Rodriguez, C.~L., Zevin, M., {Amaro-Seoane}, P., {et~al.} 2019,
  \href{http://dx.doi.org/10.1103/PhysRevD.100.043027}{\JournalTitle{Phys. Rev.
  D}, 100, 043027}

\bibitem[{Romero-Shaw {et~al.}(2020)Romero-Shaw, Lasky, Thrane, \&
  Bustillo}]{romero-shaw_gw190521_2020}
Romero-Shaw, I.~M., Lasky, P.~D., Thrane, E., \& Bustillo, J.~C. 2020,
  \href{http://arxiv.org/abs/2009.04771}{\JournalTitle{arXiv:2009.04771
  [astro-ph]}}, arXiv: 2009.04771

\bibitem[{Roupas \& Kazanas(2019)}]{roupas_generation_2019}
Roupas, Z., \& Kazanas, D. 2019,
  \href{http://dx.doi.org/10.1051/0004-6361/201937002}{\JournalTitle{A\&A},
  632, L8}

\bibitem[{Safarzadeh \& Haiman(2020)}]{safarzadeh_formation_2020}
Safarzadeh, M., \& Haiman, Z. 2020,
  \href{http://arxiv.org/abs/2009.09320}{\JournalTitle{arXiv:2009.09320
  [astro-ph, physics:gr-qc]}}, arXiv: 2009.09320

\bibitem[{{Samsing} {et~al.}(2014){Samsing}, {MacLeod}, \&
  {Ramirez-Ruiz}}]{2014ApJ...784...71S}
{Samsing}, J., {MacLeod}, M., \& {Ramirez-Ruiz}, E. 2014,
  \href{http://dx.doi.org/10.1088/0004-637X/784/1/71}{\JournalTitle{\apj}, 784,
  71}

\bibitem[{Samsing {et~al.}(2014)Samsing, MacLeod, \&
  {Ramirez-Ruiz}}]{Samsing2014}
Samsing, J., MacLeod, M., \& {Ramirez-Ruiz}, E. 2014,
  \href{http://dx.doi.org/10.1088/0004-637X/784/1/71}{\JournalTitle{ApJ}, 784,
  71}

\bibitem[{{Samsing} \& {Ramirez-Ruiz}(2017)}]{2017ApJ...840L..14S}
{Samsing}, J., \& {Ramirez-Ruiz}, E. 2017,
  \href{http://dx.doi.org/10.3847/2041-8213/aa6f0b}{\JournalTitle{\apjl}, 840,
  L14}

\bibitem[{Samsing \& {Ramirez-Ruiz}(2017)}]{Samsing2017}
Samsing, J., \& {Ramirez-Ruiz}, E. 2017,
  \href{http://dx.doi.org/10.3847/2041-8213/aa6f0b}{\JournalTitle{ApJL}, 840,
  L14}

\bibitem[{{Samsing} {et~al.}(2020){Samsing}, {Bartos}, {D'Orazio}, {Haiman},
  {Kocsis}, {Leigh}, {Liu}, {Pessah}, \& {Tagawa}}]{2020arXiv201009765S}
{Samsing}, J., {Bartos}, I., {D'Orazio}, D.~J., {et~al.} 2020,
  \JournalTitle{arXiv e-prints}, arXiv:2010.09765

\bibitem[{{Secunda} {et~al.}(2019){Secunda}, {Bellovary}, {Mac Low}, {Ford},
  {McKernan}, {Leigh}, {Lyra}, \& {S{\'a}ndor}}]{2019ApJ...878...85S}
{Secunda}, A., {Bellovary}, J., {Mac Low}, M.-M., {et~al.} 2019,
  \href{http://dx.doi.org/10.3847/1538-4357/ab20ca}{\JournalTitle{\apj}, 878,
  85}

\bibitem[{{Secunda} {et~al.}(2020){Secunda}, {Bellovary}, {Mac Low}, {Ford},
  {McKernan}, {Leigh}, {Lyra}, {S{\'a}ndor}, \& {Adorno}}]{2020ApJ...903..133S}
---. 2020,
  \href{http://dx.doi.org/10.3847/1538-4357/abbc1d}{\JournalTitle{\apj}, 903,
  133}

\bibitem[{Sesana(2016)}]{sesana_prospects_2016}
Sesana, A. 2016,
  \href{http://dx.doi.org/10.1103/PhysRevLett.116.231102}{\JournalTitle{Phys.
  Rev. Lett.}, 116, 231102}

\bibitem[{{Seth} {et~al.}(2008){Seth}, {Blum}, {Bastian}, {Caldwell}, \&
  {Debattista}}]{2008ApJ...687..997S}
{Seth}, A.~C., {Blum}, R.~D., {Bastian}, N., {Caldwell}, N., \& {Debattista},
  V.~P. 2008, \href{http://dx.doi.org/10.1086/591935}{\JournalTitle{\apj}, 687,
  997}

\bibitem[{{Silsbee} \& {Tremaine}(2017)}]{2017ApJ...836...39S}
{Silsbee}, K., \& {Tremaine}, S. 2017,
  \href{http://dx.doi.org/10.3847/1538-4357/aa5729}{\JournalTitle{\apj}, 836,
  39}

\bibitem[{{Sz{\"o}lgy{\'e}n} \& {Kocsis}(2018)}]{2018PhRvL.121j1101S}
{Sz{\"o}lgy{\'e}n}, {\'A}., \& {Kocsis}, B. 2018,
  \href{http://dx.doi.org/10.1103/PhysRevLett.121.101101}{\JournalTitle{\prl},
  121, 101101}

\bibitem[{Tagawa {et~al.}(2020)Tagawa, Kocsis, Haiman, Bartos, Omukai, \&
  Samsing}]{tagawa_eccentric_2020}
Tagawa, H., Kocsis, B., Haiman, Z., {et~al.} 2020,
  \href{http://arxiv.org/abs/2010.10526}{\JournalTitle{arXiv:2010.10526
  [astro-ph]}}

\bibitem[{Tanikawa {et~al.}(2020)Tanikawa, Kinugawa, Yoshida, Hijikawa, \&
  Umeda}]{tanikawa_population_2020}
Tanikawa, A., Kinugawa, T., Yoshida, T., Hijikawa, K., \& Umeda, H. 2020,
  \href{http://adsabs.harvard.edu/abs/2020arXiv201007616T}{\JournalTitle{arXiv
  e-prints}, 2010, arXiv:2010.07616}

\bibitem[{Toubiana {et~al.}(2020)Toubiana, Sberna, Caputo, Cusin, Marsat, Jani,
  Babak, Barausse, Caprini, Pani, Sesana, \&
  Tamanini}]{toubiana_detectable_2020}
Toubiana, A., Sberna, L., Caputo, A., {et~al.} 2020,
  \href{http://adsabs.harvard.edu/abs/2020arXiv201006056T}{\JournalTitle{arXiv
  e-prints}, 2010, arXiv:2010.06056}

\bibitem[{Vallisneri(2008)}]{Vallisneri2007}
Vallisneri, M. 2008,
  \href{http://dx.doi.org/10.1103/PhysRevD.77.042001}{\JournalTitle{Physical
  Review D}, 77, 042001}

\bibitem[{{VanLandingham} {et~al.}(2016){VanLandingham}, {Miller}, {Hamilton},
  \& {Richardson}}]{2016ApJ...828...77V}
{VanLandingham}, J.~H., {Miller}, M.~C., {Hamilton}, D.~P., \& {Richardson},
  D.~C. 2016,
  \href{http://dx.doi.org/10.3847/0004-637X/828/2/77}{\JournalTitle{\apj}, 828,
  77}

\bibitem[{Virtanen {et~al.}(2020)Virtanen, Gommers, Oliphant, Haberland, Reddy,
  Cournapeau, Burovski, Peterson, {Weckesser}, {Bright}, {van der Walt},
  {Brett}, {Wilson}, {Jarrod Millman}, {Mayorov}, {Nelson}, {Jones}, {Kern},
  {Larson}, {Carey}, {Polat}, {Feng}, {Moore}, {Vand erPlas}, {Laxalde},
  {Perktold}, {Cimrman}, {Henriksen}, {Quintero}, {Harris}, {Archibald},
  {Ribeiro}, {Pedregosa}, {van Mulbregt}, \& {Contributors}}]{2020SciPy}
Virtanen, P., Gommers, R., Oliphant, T.~E., {et~al.} 2020, \JournalTitle{Nature
  Methods}

\bibitem[{{Walcher} {et~al.}(2005){Walcher}, {van der Marel}, {McLaughlin},
  {Rix}, {B{\"o}ker}, {H{\"a}ring}, {Ho}, {Sarzi}, \&
  {Shields}}]{2005ApJ...618..237W}
{Walcher}, C.~J., {van der Marel}, R.~P., {McLaughlin}, D., {et~al.} 2005,
  \href{http://dx.doi.org/10.1086/425977}{\JournalTitle{\apj}, 618, 237}

\bibitem[{Walt {et~al.}(2011)Walt, Colbert, \& Varoquaux}]{walt_numpy_2011}
Walt, S. v.~d., Colbert, S.~C., \& Varoquaux, G. 2011,
  \href{http://dx.doi.org/10.1109/MCSE.2011.37}{\JournalTitle{Computing in
  Science Engineering}, 13, 22}

\bibitem[{Wen(2003)}]{wen_eccentricity_2003}
Wen, L. 2003, \href{http://dx.doi.org/10.1086/378794}{\JournalTitle{ApJ}, 598,
  419}

\bibitem[{{Zevin} {et~al.}(2019){Zevin}, {Samsing}, {Rodriguez}, {Haster}, \&
  {Ramirez-Ruiz}}]{2019ApJ...871...91Z}
{Zevin}, M., {Samsing}, J., {Rodriguez}, C., {Haster}, C.-J., \&
  {Ramirez-Ruiz}, E. 2019,
  \href{http://dx.doi.org/10.3847/1538-4357/aaf6ec}{\JournalTitle{\apj}, 871,
  91}

\bibitem[{Zhan {et~al.}(2019)Zhan, Wang, Ni, Gao, Wang, He, Li, Zhou, Chen,
  Zhong, Tang, Yao, Zhu, Xiong, Lu, Yu, Cheng, Liu, Liang, Xu, He, Ke, Tan, \&
  Luo}]{zhan_zaiga_2019}
Zhan, M.-S., Wang, J., Ni, W.-T., {et~al.} 2019,
  \href{http://dx.doi.org/10.1142/S0218271819400054}{\JournalTitle{Int. J. Mod.
  Phys. D}, 29, 1940005}

\bibitem[{Zwick {et~al.}(2020)Zwick, Capelo, Bortolas, Mayer, \&
  Amaro-Seoane}]{zwick_improved_2020}
Zwick, L., Capelo, P.~R., Bortolas, E., Mayer, L., \& Amaro-Seoane, P. 2020,
  \href{http://dx.doi.org/10.1093/mnras/staa1314}{\JournalTitle{MNRAS}, 495,
  2321}

\end{thebibliography}
\appendix
\section{Signal-to-noise ratio and Fisher Analysis} \label{app:a}
The total signal-to-noise ratio (SNR) of a GW signal in each detector is estimated as a sum of the SNRs of each individual harmonic

\begin{equation}
{\rm SNR}_{i}^2 \approx 2 \displaystyle\sum_{n=1}^{N} \int_{f_a}^{f_b} \frac{Q h_{{\rm c},n}^2 (f)}{f {\cal S}_i(f)} \frac{{\rm d}f}{f} \ ,
\label{eqn:snr}
\end{equation}
where ${\cal S}_i (f)$ is the power spectral density of the $i$th detector, $Q$ is a factor associated with the source orientation and detector antenna pattern, which we assume to be optimal, and the characteristic strain at the $n$th harmonic is
\begin{equation}
h_{{\rm c},n} (f) = \frac{1}{\pi D} \sqrt{\frac{2 G}{c^3} \frac{{\rm d}E_n}{{\rm d}f_{\rm r}}}\ .
\label{eqn:charstrain}
\end{equation}
\noindent and the energy emitted per GW frequency at the $n$th harmonic is \citep[e.g.,][]{huerta_detection_2015,dorazio_black_2018}
\begin{equation}
\frac{{\rm d}E_n}{{\rm d} f_{\rm r}} = \frac{G^{2/3}\pi^{2/3} {\cal M}^{5/3}}{3 (1+z)^{1/3} f^{1/3}} \left(\frac{2}{n} \right)^{2/3} \frac{g(n,e)}{F(e)} = \frac{ (2\pi)^{2/3}}{3} \frac{G^{2/3} {\cal M}^{5/3}}{f_{\rm orb}^{1/3}} \frac{g (n,e)}{n {\cdot}F(e)}\ .
\end{equation}

We can further use equation \eqref{eqn:charstrain} for computing our Fisher matrix analysis of the eccentricity.  For GW measurement uncertainties, the Fisher information matrix (FIM) can be expressed using a similar ``overlap integral'' to that used to calculate the SNR in equation \eqref{eqn:snr}.  Specifically, the $i$th and $j$th element of the FIM is \citep[e.g.,][]{Finn1996}

\begin{equation}
F_{ij} = \sum_{n=1}^N \left< \frac{\partial h_n}{\partial \theta^i} ~\vline~ \frac{\partial h_n}{\partial \theta^j}\right>,
\label{eqn:fim}
\end{equation}

\noindent where $h_{n,i}$ is the partial derivative, $\frac{\partial h_n}{\partial \theta^i}$, of the frequency-domain waveform for the $n$th harmonic with respect to the $i$th parameter of our waveform, and the $\left< ~|~ \right>$ notation indicates an overlap integral of the form 

\begin{equation}
\left< a| b \right> \equiv \ 4 \Re \int_0^{\infty} \frac{a(f) b^*(f)}{S_i(f)} \ {\rm d}f \ .
\end{equation}

\noindent For this analysis, we consider a 4-dimensional parameter space $\mathbf{\theta} = \{M, \nu, e, D\}$, consisting of the total mass, symmetric mass ratio, eccentricity, and luminosity distance, respectively.

It can be shown \cite[e.g.,][]{Vallisneri2007} that with sufficiently high SNR, the uncertainties for GW parameter estimation in idealized Gaussian noise are themselves given by multidimensional Gaussians of the form 

\begin{equation}
p(\mathbf{\theta}| s) \propto p(\mathbf{\theta}) \exp\left[-\frac{1}{2}F_{ij}\Delta \theta^i \Delta \theta^j\right]
\label{eqn:prob}
\end{equation}

\noindent where $\Delta \theta^i$ is the separation between the $i$th parameter and the maximum likelihood value, and $p(\mathbf{\theta})$ is the prior probability distribution on the parameters $\mathbf{\theta}$ (which we assume to be uniform for this analysis).  Note that \eqref{eqn:prob} can be interpreted in either a frequentist framework (where it corresponds to the Cram\'er-Rao bound on any unbiased estimator of the GW source parameters) or a Bayesian framework (where it corresponds to the covariance of the posterior probability about the true source parameters, assuming the prior to be constant over that range); however, both interpretations yield the same results \citep{Vallisneri2007}.  

The uncertainties on our measured eccentricities that we show in Panel D of \ref{fig:hc_snr} are calculated using Equation \eqref{eqn:fim}, with the same noise curve and waveforms described in that section.  We use the characteristic strains from \eqref{eqn:charstrain} as our GW template, and calculate the uncertainties and correlations between our parameters as

\begin{align}
\sigma_{i} &= \sqrt{\Sigma^{ii}} \\
\sigma_{ij} &= \frac{\Sigma^{ij}}{\sqrt{\Sigma^{ii}\Sigma^{jj}}}\nonumber
\end{align}

\noindent where $\Sigma^{ij} = (F^{-1})^{ij}$ is the inverse of the FIM.  Note that we calculate the full 4-dimensional FIM, but only report the uncertainties on $e$ in the main text.  
\end{document}